# Validation of SPAMM Tagged MRI Based Measurement of 3D Soft Tissue Deformation

*Kevin M. Moerman [a),b)], Andre M. J. Sprengers [b)], Ciaran K. Simms [a)], Rolf M. Lamerichs [b),c)], Jaap Stoker [b)] and Aart J. Nederveen [b)]*

[a)] Trinity Centre for Bioengineering, School of Engineering, Parsons Building, Trinity College, Dublin 2, Ireland
[b)] Radiology Department, Academic Medical Centre, Meibergdreef 9, 1105 AZ Amsterdam, The Netherlands
[c)] Philips Research, High Tech Campus 5, 5656 AE Eindhoven, The Netherlands

## Abstract

**Purpose:** This study presents and validates a novel (non-ECG-triggered) MRI sequence based on SPAtial Modulation of the Magnetization (SPAMM) to non-invasively measure 3D (quasi-static) soft tissue deformations using only six acquisitions (three static and three indentations). In current SPAMM tagged MRI approaches data is typically constructed from many repeated motion cycles. This has so far restricted its application to the measurement of highly repeatable and periodic movements (e.g. cardiac deformation). In biomechanical applications where soft tissue deformation is artificially induced, often by indentation, significant repeatability constraints exist and, for clinical applications, discomfort and health issues generally preclude a large number of repetitions.

**Methods:** A novel (non-ECG-triggered) SPAMM tagged MRI sequence is presented whereby a single 1-1 (first order) SPAMM set is acquired following a 3D Transient Field Echo acquisition. Full 3D deformation measurement is achieved through the combination of only six acquisitions (three static and three motion cycles). The 3D deformation measurements were validated using quasi-static indentation tests and marker tracking in a silicone gel soft tissue phantom. In addition, the technique's ability to measure 3D soft tissue deformation *in-vivo* was evaluated using indentation of the biceps region of the upper arm in a volunteer.

**Results:** Following comparison to marker tracking in the silicone gel phantom, the SPAMM tagged MRI based displacement measurement demonstrated sub-voxel accuracy with a mean displacement difference of 72 μm and a standard deviation of 289 μm. In addition, precision of displacement magnitude was evaluated for both the phantom and volunteer data. The standard deviations of the displacement magnitude with respect to the average displacement magnitude were 75 μm and 169 μm for the phantom and volunteer data respectively.

**Conclusions:** The sub-voxel accuracy and precision demonstrated in the phantom in combination with the precision comparison between the phantom and volunteer data provide confidence in the methods presented for measurement of soft tissue deformation *in vivo*. To our knowledge, since only six acquisitions are required, the presented methodology is the fastest SPAMM tagged MRI method currently available for the non-invasive measurement of quasi-static 3D soft tissue deformation.

# I. INTRODUCTION

Magnetic resonance imaging (MRI) based analysis of human soft tissue deformation *in-vivo* has many important applications, for example the study of cardiac biomechanics[1, 2], assessment of tumor motion[3] and preoperative planning[4]. Furthermore, when combined with inverse analysis, it enables the non-invasive determination of the mechanical properties of human soft tissue[5]. This has potential for tumor detection[6] and the development of constitutive models of human soft tissue. These constitutive formulations are vital for computational models of tissue stress and strain, muscle load and joint reaction force, all of which cannot generally be measured *in vivo*[7, 8]. Such models are applied for example in the prediction of tissue trauma[9], pressure ulcers[10, 11], analysis of tissue engineering constructs[12] and realistic surgical simulation[13]. The work presented here is part of a larger study using indentation tests on the human arm and MRI based measurement of soft tissue deformation to determine the mechanical properties of passive living human skeletal muscle tissue.

MRI provides excellent soft tissue contrast without exposing subjects to ionizing radiation. As such, a large variety of MRI based methods have been developed to non-invasively analyze soft tissue deformation. These can be roughly subdivided into; *1) methods that rely on anatomical features* (e.g. using correlation methods[14, 15] or deformable models and non-rigid image registration[16]) and, since anatomical features may be insufficient, *2) methods that rely on implanted markers*[17], and finally *3) methods that rely on specialized MRI sequences and signal modulation*, where phase contrast MRI methods[18, 19] or methods based on SPAtial Modulation of the Magnetization (SPAMM)[20, 21] are common[1, 2, 22, 23]. The latter is the focus of this paper.

In SPAMM tagged MRI sequences the magnetization is modulated using radiofrequency pulses and magnetic field gradients, resulting in saturated bands in the magnetization distribution and, as a consequence, contrasting patterns in the image data. These patterns act as temporary markers locked in the tissue whose appearance reflects any underlying tissue motion. Typically SPAMM tagged MRI methods are gated and image acquisition is synchronized with the motion cycle (e.g. to heartbeats using electrocardiograms), whereby only part of the k-space is acquired within each cycle, and acquisitions from multiple motion cycles are used to compose a dataset representing a single motion cycle. This has confined the application of SPAMM tagged MRI to the analysis of highly repeatable and periodic movements and has mainly been applied to the heart[1, 2, 24], but more recently also to study repeated (natural or induced) motions of the tongue[25], brain[26], eyes[27] and (rat) skeletal muscles[28]. The number of repetitions used varies depending on the temporal and spatial resolution from 16 per slice[25] to over a hundred[26]. In the latter study 2D brain strain was estimated following 144 repeated angular accelerations of the head. However, discomfort, health issues and clinical conditions which affect repeatability (e.g. cardiac arrhythmia) often preclude the use of a large number of repetitions. Faster approaches have recently been developed for estimation of 2D lung tissue movement[29] but these cannot be applied to other tissue sites or to a clinical setting as inhaled hyperpolarized $^{3}He$ gas was used as a contrast medium, thus accelerating the image acquisition.

As the above review demonstrates current SPAMM tagged MRI methods often require a large number of repetitions. This is however undesirable for biomechanical application such as muscle indentation since it: *1) causes volunteer discomfort, 2) results in long scanning times, 3) hinders the analysis of dynamic biomechanical tissue responses e.g. visco-elasticity and pre-conditioning and, 4) places repeatability constraints on both the imaging and load inducing devices.* Hence the purpose of the current study was to develop and validate SPAMM tagged MRI based methods to accurately measure 3D quasi-static soft tissue deformation requiring a minimum of repeated motion (indentation) cycles. The focus of this paper is therefore on optimization for speed and on the reduction of the number of repeated acquisitions required.

Many approaches have been proposed to derive motion from SPAMM tagged MRI data; e.g. using non-rigid image registration[30], computational models[31], harmonic phase (HARP) methods[32] and 3D tag surface analysis[33, 34] (for detailed reviews on this topic see[1, 2, 22]). In general, post-processing methods aim at deriving tissue displacement via the analysis of the change of shape of tags or derived phase features. However factors such as noise, artifacts and anatomical disturbances may cause ambiguities in tag analysis and this has led to the use of various regularization strategies. For example, large numbers of repetitions, signal averages and interpolation have been employed, as well as assumptions about the nature, repeatability and smoothness of the deformation[2] (e.g. by fitting prolate spheroidal B-spline models of the heart to multiple 1D displacement measurements[32]). Since the validation of constitutive models of soft tissues requires accurate measurement of tissue anisotropy and non-linearity, assumptions about the nature of the deformation are not desirable in the current study. In contrast, if intersections are computed for tag surfaces in three initially orthogonal directions, these provide material points[33, 34] which, when tracked, offer a reliable measurement of 3D displacement without the requirement for assumptions on the nature of the deformation. Therefore this approach was adopted for the current study. Many current post-processing methods have been developed for data derived from repeated acquisitions which provides increased resolution and signal to noise ratio (SNR) with respect to the data used in the current study. Therefore a novel post-processing framework, based on tracking of tag surface intersections, was developed. This allows for the accurate computation of 3D soft tissue deformation from low resolution and noisy data while avoiding unrealistic assumptions on the nature of the deformation.

Validation of the derived displacement measurements requires comparison against a "gold standard" reference measure. Since such a reference measure is often not present *in-vivo*, many studies have used numerical and physical models for validation. However in many cases the reliability of the reference itself is not known or evaluated. In one study validation with implanted crystals and sonomicrometric measurements was performed[35], but crystal locations were verified manually and matching problems between MR and sonomicrometric measurements occurred. Recently a novel validation framework for MRI based deformation measurement was proposed based on a deformable silicone gel phantom containing contrasting spherical markers[36]. The "gold standard" reference is made available through marker tracking. The validation methodology was itself independently validated using image simulations (error smaller than 0.12 voxels) and therefore provides a reliable "gold standard" for the evaluation of the SPAMM derived deformation measurement. This validation approach was therefore adopted for the current study.

This paper presents a novel (non-ECG-triggered) SPAMM tagged MRI sequence for the accurate measurement of (quasi-static) 3D soft tissue deformation. A parallel tag pattern is introduced using a single 1-1 (first order) SPAMM pre-pulse, followed by a 3D transient field echo (TFE) acquisition. In order to derive 3D deformation the intersections of three sets of orthogonally placed tag surfaces are computed and tracked from their initial to their deformed configurations. The tag surfaces are segmented using a newly developed sheet marching algorithm. Only six acquisitions (three static and three indentations) are required for measurement of 3D soft tissue deformation. To the best of our knowledge this makes the SPAMM tagged MRI based techniques presented here the fastest available for the non-invasive measurement of (quasi-static) 3D soft tissue deformation. The deformation measurements were validated by indentation tests (using a custom designed MRI compatible soft tissue indentor) and marker tracking[36] in a silicone gel phantom. In addition the performance of the method was evaluated *in-vivo* using indentation of the biceps region of the upper arm of a volunteer. Marker tracking in the phantom allowed for the determination of the accuracy and precision of the displacement measurements, and the latter was also compared to the measured precision *in-vivo*.

# II. METHODS

## II.A. MRI sequence design

For the current study a novel MRI sequence, based on SPAMM[20, 21] was used which forms an expansion of a sequence earlier developed for 1D bowel motion estimation[37] to the measurement of 3D soft tissue deformation. The sequence was applied to a silicone gel phantom for validation and to a volunteer for *in-vivo* evaluation. A schematic for the pulse sequence design is shown in Fig. 1 and Table I provides a summary of the scanning parameters and configurations used in this study and illustrates three slices for each data set recorded. A 5 ms single 1-1 (first order) SPAMM pre-pulse imposes a (Fig. 1. parts A) temporary sinusoidal modulation on the Z-magnetization and thus also on the signal magnitude causing it to vary sinusoidally from normal to severely reduced. The reduced signal regions form a parallel surface pattern in the tissue (visible as low intensity lines in 2D slices). These surfaces are generally referred to as tags. After the tag pre-pulse a time delay (Fig. 1. parts B) is introduced during which the tissue indentation occurs. Subsequently, after the indentation is complete and the time delay has ended, a full 3D volume is acquired using a single 3D Transient Field Echo (TFE) read-out. A TFE read-out (equivalent to Turbo Field Echo in Philips nomenclature) was chosen in order to minimize acquisition times. The 3D TFE sequence was configured with a Cartesian acquisition mode in k-space, the profile order was set to low-high, a radial turbo direction was used, and in the z-direction the resolution was doubled during reconstruction. No acceleration techniques were used. A tag spacing of 6 mm and 9 mm was used for the phantom and human volunteer respectively. The sequence was implemented on a 3.0 Tesla Philips Intera scanner (Philips Healthcare, Best, The Netherlands), scans were performed using two FLEX-M coils. A total of six acquisitions are used for computation of tissue deformation, three repeated data sets with mutually orthogonal SPAMM directions for both the initial and deformed configuration.

Since the tag surface pattern is temporarily locked in the tissue (fades due to $T_1$ relaxation), the deformation of the tissue, that occurred between pre-pulse and readout, is reflected in the deformation of the tag surface pattern. However unless the assumption is made that tag-surfaces are planar in their un-deformed configuration (and if no within-surface reference points are available), a single set of tag surfaces only provides tag surface shape, not (a component of) displacement. Since field in-homogeneities at the region of interest (upper-arm) cause the initial tag surfaces to be mildly deformed the initial states were not assumed flat. Therefore for both the initial and deformed configuration three repeated data sets were acquired with mutually orthogonal SPAMM directions. The intersection points of the orthogonal tag surfaces provide trackable material points and the 3D displacement of these material points is given as the difference between their coordinates in the initial and deformed configuration. Due to the tag spacing applied a grid of 6x6x6 mm and 9x9x9 mm of points can be tracked for the phantom and volunteer respectively.

In many SPAMM tagged MRI studies acquisitions are gated for instance in cardiac applications[2] by triggering the scan sequence to the electrocardiogram in order to synchronize imaging with the heart motion cycle. However in the current study external loading (indentation) is applied to induce deformation and no physiological trigger is available. As such the external trigger dependence was removed. Instead the scanner was used to generate a 40 ms Transistor-Transistor Logic (TTL) pulse allowing for the appropriate and repeatable timing of the indentation. In Fig.1. the *n* stands for number of dynamics. For validation purposes 20 dynamics were acquired for each SPAMM direction. During the acquisition of these dynamics the indentor was repeatedly activated. Prior to each indentation an initial, un-deformed configuration was acquired during which the tissue/gel was static. The indentor was then activated following a TTL pulse generated after SPAMM deposition (Fig. 1. parts A). Then, after the indentor is static, in its final location and after the delay time has been reached the deformed configuration is acquired using the 3D TFE read-out (Fig. 1. parts C). No motion occurs during this read-out. The indentor then is slowly moved to its original position following a pause of 3s. The dynamics acquired during the pause and slow un-loading were not used. Once the indentor has returned to its initial location the above cycle is repeated and a new initial configuration is acquired. Scanning parameters were optimized for each scan direction, leading to varying scan durations and thus the varying numbers of acquired initial and deformed configurations for each scan direction shown in Table I.

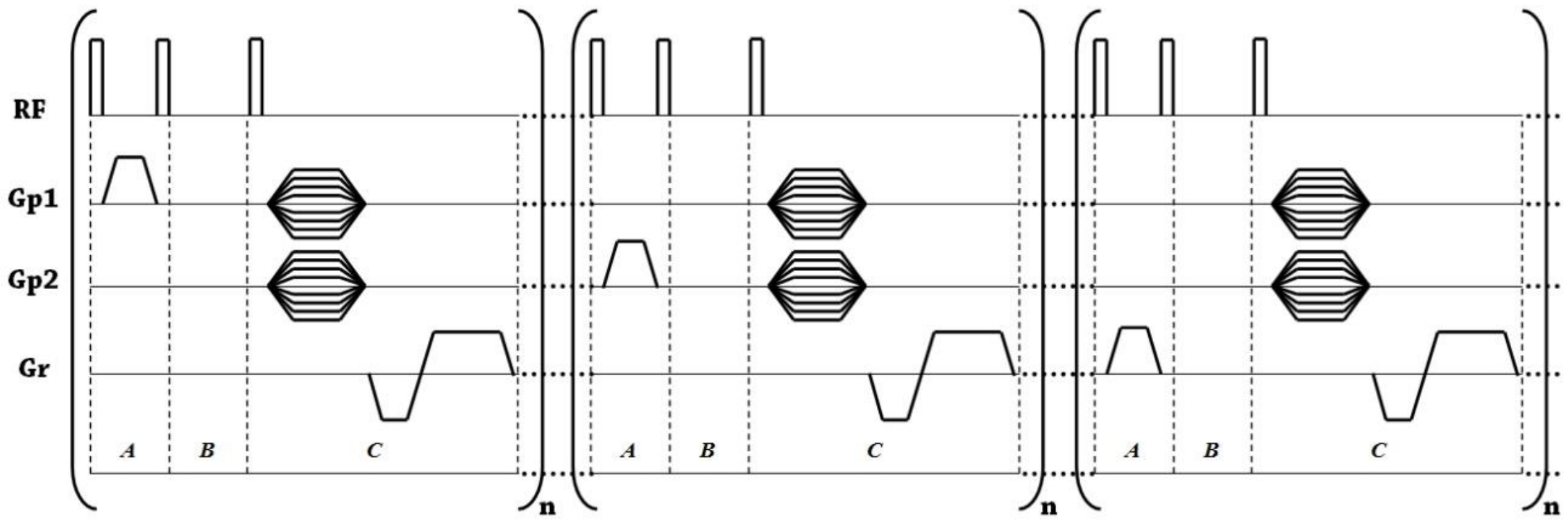

*FIG. 1. Diagram of the SPAMM pulse sequence and 3D TFE read-out. The 1-1 SPAMM pre-pulse (A) modulates the signal followed by a desired time delay (B) after which a 3D volume is acquired (C). The acquisition can be repeated n times for each direction.*

*TABLE I. MRI acquisition matrix*

| Initial configuration | Deformed configuration | Number of acquisitions | SPAMM type, Slice orientation, $T_R/T_E$ (ms) | Field of view (mm), Acquisition matrix (frequency x phase), Read-out time (ms) | # slices | Reconstructedvoxel dimensions (mm) |
|---|---|---|---|---|---|---|
| Volunteer SPAMM tagged MRI | | | | | | |
| 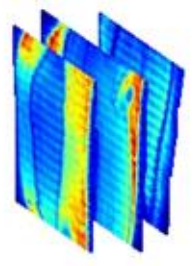 | 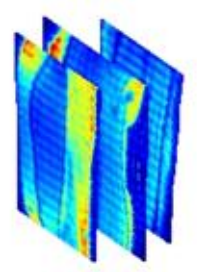 | 6 Initial<br>7 Deformed | Z 1-1 SPAMM<br>YZ slices<br>2.18/1.02 | 130x130x52.5<br>64x65<br>1800 | 35 | 1.35x1.35x1.5 |
| 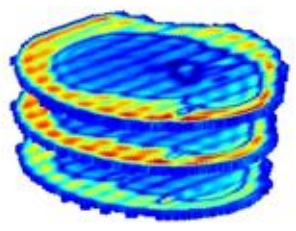 | 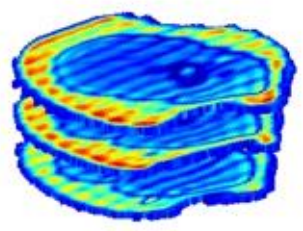 | 4 Initial<br>4 Deformed | Y 1-1 SPAMM<br>XY slices<br>2.29/1.11 | 253x253x45<br>124x84<br>1500 | 15 | 0.88x0.88x3 |
| 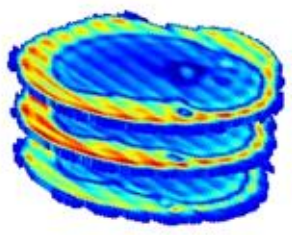 | 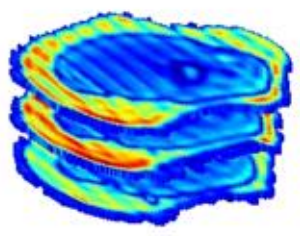 | 4 Initial<br>4 Deformed | X 1-1 SPAMM<br>XY slices<br>2.18/1.02 | 253x253x45<br>124x84<br>1500 | 15 | 0.88x0.88x3 |
| Phantom SPAMM tagged MRI | | | | | | |
| 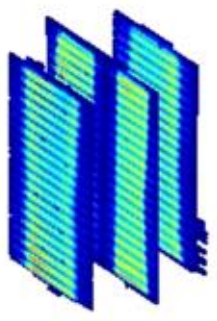 | 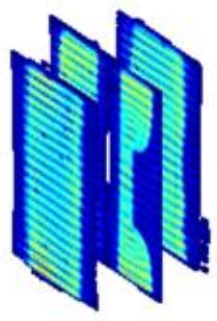 | 10 Initial<br>5 Deformed | Z 1-1 SPAMM<br>YZ slices<br>2.39/1.16 | 130x130x67.5<br>88x84<br>2100 | 45 | 1.35x1.35x1.5 |
| 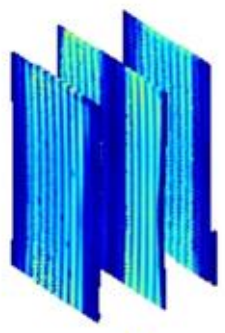 | 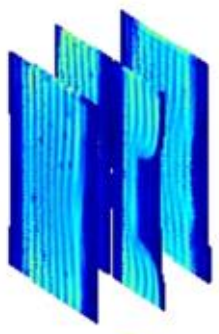 | 10 Initial<br>5 Deformed | Y 1-1 SPAMM<br>YZ slice<br>2.51/1.21 | 130x130x52.5<br>80x80<br>2400 | 35 | 1.35x1.35x1.5 |
| 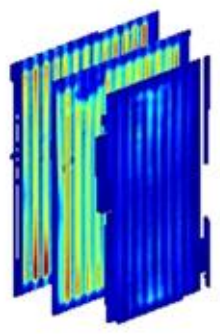 | 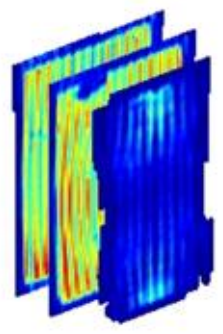 | 7 Initial<br>6 Deformed | X 1-1 SPAMM<br>XZ slices<br>2.39/1.16 | 130x130x67.5<br>88x87<br>2100 | 45 | 1.25x1.25x1.5 |
| Phantom T2 weighted MRI | | 1 Initial<br>1 Deformed | T2 Weighted<br>XZ slices<br>2500/637.79 | 120x120x80<br>240x240<br>10min | 160 | 0.47x0.47x0.5 |
| 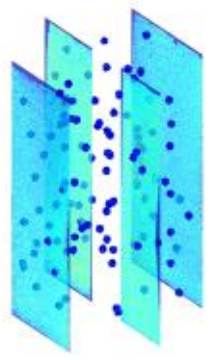 | 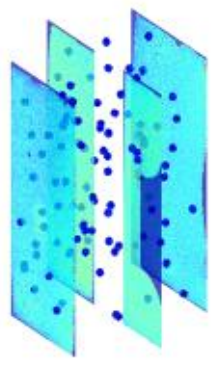 | | | | | |

## II.B. The validation set-up: indentor and soft tissue phantom

Validation of the soft tissue deformation measurement was achieved using indentation tests on a silicone gel phantom (Fig. 2A). The phantom's cylindrical shape (120 mm long and 80 mm in diameter) and stiff poly-oxy-methylene (POM) "bone-like" core (20mm in diameter) represents an idealized geometry of the upper arm, allowing simulation of comparable deformation modes. The silicone gel (SYLGARD® 527 A&B Dow Corning, MI, USA) simulates soft tissue as its stiffness[5] and MRI properties[38] are within the range of human soft tissue. For human muscle tissue at 3 T the literature reports $T_1$ and $T_2$ relaxation times in the range of 898~1420 ms and 29~50 ms respectively [39-41]. Estimates for the silicone gel used for the current study (gel mixing ratio 1:1), derived from regions of interest in an inversion recovery and a multi echo sequence, are 1026ms (standard deviation 23 ms) and 104 ms (standard deviation 2 ms) respectively. The similarity of the $T_1$ relaxation times is most relevant to the current study since it relates to SPAMM tag persistence.

In order to create the "gold standard" reference for measurement of tissue deformation spherical POM balls of 3±0.05 mm diameter (The Precision Plastic Ball Co Ltd, Addingham, UK) were embedded within the gel volume at approximately 15 mm intervals. These markers have low signal compared to the gel and can be tracked from high resolution (0.5mm isotropic voxels) $T_2$ weighted images to provide an independent measure of deformation thus allowing for the evaluation of accuracy of the deformation measurement. The bottom row in Table I shows three slices for the $T_2$ weighted data in the initial and deformed configuration. In addition, the segmented markers are shown here as voxels. The marker tracking methods have been previously described[36] and demonstrated errors under 0.12 voxels (0.05 mm).

In addition to the phantom validation the ability to measure soft tissue deformation *in-vivo* was demonstrated using indentation tests on the biceps region of the upper arm (see Fig. 2B) of a healthy volunteer (female, age 24, height 1.65 m, weight 65 kg, ethical approval and informed consent obtained).

The gel phantom and the volunteer's upper arm were subjected to static transverse ramp (~12 mm) and hold (3 s) indentations (at a speed of ~40 mm/s) using a custom designed, hydraulically powered and MRI compatible soft tissue indentor (Fig. 2B). The indentor is also equipped with a high speed MRI compatible force sensor enabling measurement of force boundary conditions required for inverse analysis of tissue mechanical properties. The indentor head is cylindrical (45 mm in diameter) and its speed and depth can be varied via a computer controlled hydraulic master cylinder. As mentioned before appropriate and repeatable timing of the indentation was achieved via triggering towards a 40 ms TTL pulse generated by the scanner prior to imaging.

Fig 2C and 2D illustrate several image slices with the tag-lines for the phantom and volunteer data in the initial and deformed configuration. In addition iso-surfaces are shown which clearly indicate the circular indentation sites in the deformed configuration. For clarity all figures are presented with a similar 3D orientation such that the reader maintains a reference towards the orientation and location of the indentation.

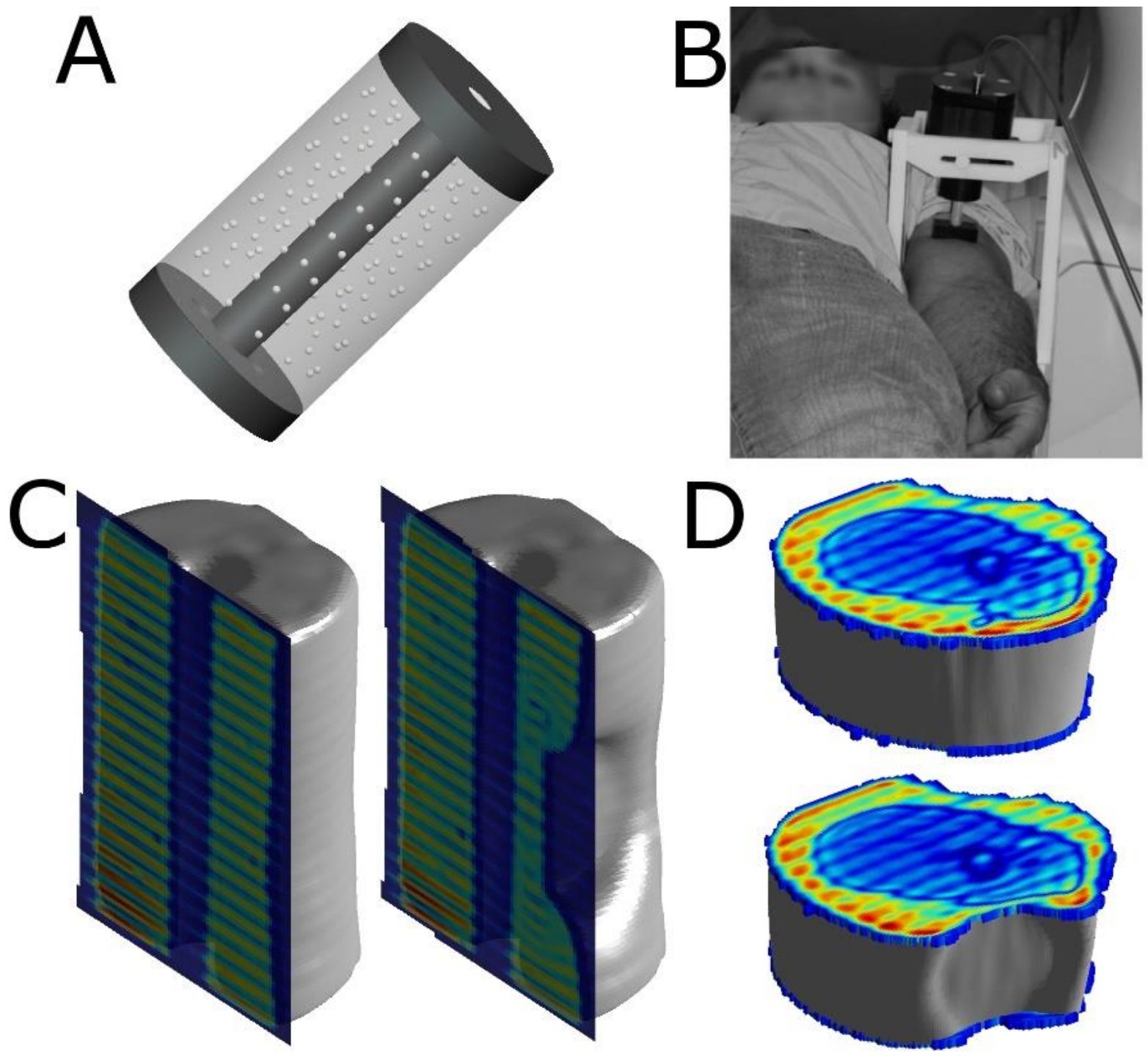

*FIG. 2. The silicone gel soft tissue phantom (A), the MRI compatible indentor placed at volunteer's upper arm (B), iso-surface plot showing the indentation sites for the phantom (C) and volunteer (D).*

## II.C. Deriving tissue deformation from the SPAMM tagged MRI data

In order to extract soft tissue deformation from the SPAMM tagged MRI data was analyzed using novel post-processing methods developed in Matlab (The Mathworks Inc., Natick, MA, USA). The MRI data (DICOM files) were uploaded and slices were combined into 3D volume matrices and the data was normalized towards maximum intensity. No filtering was applied. Post-processing was performed in the following steps: *A) Logic masking*, to identify potential tag-voxels, *B) Connectivity analysis*, to remove irregularities (e.g. anatomical disturbances), *C) Sheet marching* to segment the tags and fit surfaces, and finally *D) Calculation of tag surface intersections*, whereby tag surface intersection points for the initial and deformed configuration were used to derive the displacement vector field. The latter two geometric operations (C and D) occur in a regular Cartesian space (patient coordinate system) whereby the non-uniform voxel dimensions are taken into account.

### *II.C.1. Logic masking*

Logic masking was used to determine the tag surface orientation and to identify potential tag voxels. A 3D cross-shaped mask was used to act as a logic operator testing the mask environment for certain criteria. The mask dimensions can be altered according to the tag thickness. The logic masking checks whether the outer mask elements are of higher intensity than the voxel it is centered on. The cross-shaped mask can be seen as being composed of three mutually orthogonal segments acting in the nominal axis directions. The overall dominance of any one of these segments allows for the detection of global tag surface orientation. For voxels that are part of a tag surface there is at least one higher intensity element on both sides of one of the orthogonal segments. The result of this analysis step was a binary logic matrix. Fig. 3A shows Y 1-1 SPAMM data for the human volunteer. In Fig. 3B the tag voxels for these slices, identified using masking are overlaid as voxels (gray and black). However MRI is complicated by noise, possible artifacts and, in addition anatomical features may induce additional intensity variations *in-vivo*. Such disturbances may result in false positives tag voxels (e.g. at the muscle/fat boundary in Fig. 3). However tag surfaces are by definition continuous structures (in the absence of shearing) and this knowledge can be used *a priori* to correct for possible false positive tag voxels using the connectivity analysis presented next.

### *II.C.2. Connectivity analysis*

In order to remove false positive tag voxels connectivity analysis was performed. A voxel's tag connectivity in a certain direction was defined as the number of tag voxels it is connected to in this direction. For instance each tag voxel in a column of *n* tag voxels has a connectivity of *n* in the column direction. If a voxel is part of a continuous tag its connectivity in the perpendicular direction should reflect the thickness of the tag, while in the parallel directions it should reflect the local tag width and length respectively. These connectivity measures distinguish false positives from tag voxels, and allow for their removal. Voxels discarded in this way, e.g. those near the bone and the muscle/fat boundary, are shown in gray in Fig. 3B. For the phantom data, tag voxels that formed bridges between two adjacent tags (due to markers) were removed in this way.

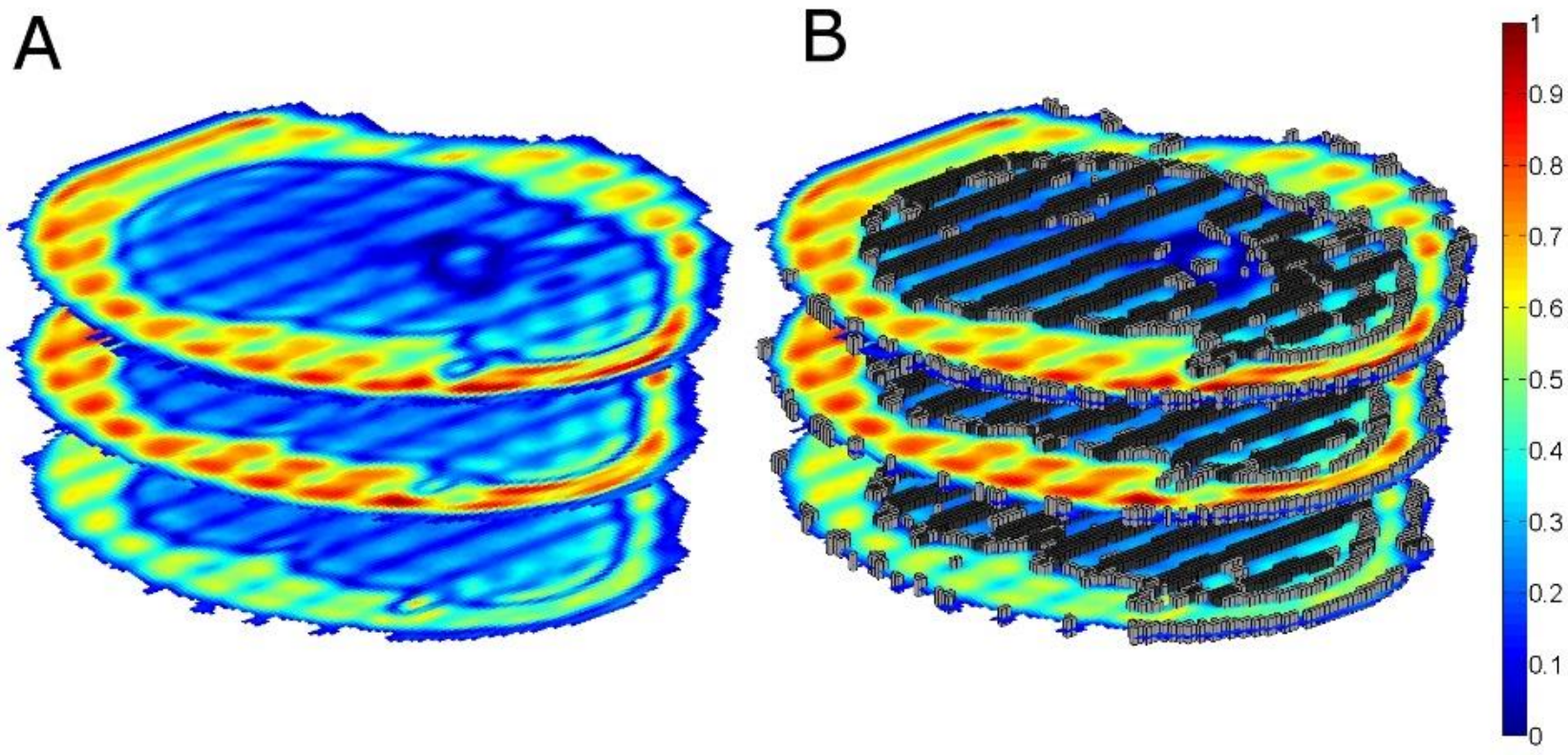

*FIG. 3. Three slices for the human Y 1-1 SPAMM data (A) and the same slices with overlaid tag-voxels, those that pass the connectivity analysis are shown in black while those that are discarded are shown in gray (B).*

### *II.C.3. Sheet marching*

Segmentation of the tag surfaces was achieved using a novel sheet-marching algorithm in which tag surfaces were represented by cubic spline surfaces that "grow" using two moving fronts from manually determined start and end locations (graphical user interface assisted selection involving around 5 minutes of user interaction per data set). Each surface grows from the extremities inwards and marches towards the opposite side guided by the tag voxels identified using masking and connectivity analysis. This process consists of two main steps which are repeated until the surface was complete: *1) Extrapolation of surface fronts, 2) Update extrapolated fronts using weighted averaging.*

II.C.3.a. Extrapolation of surface fronts

Tag surfaces were represented as cubic spline surfaces which were determined through minimization of the following expression (see Matlab *csaps* function and[42]):

$$\mu\sum_{i=1}^{n}\left(g\left(\mathbf{x}_i\right)-f\left(\mathbf{x}_i\right)\right)^2+\left(1-\mu\right)\int_{x_1}^{x_n}\left(\frac{d^2f}{d\mathbf{x}^2}\right)^2 d\mathbf{x} \qquad 1$$

Here $g\left(\mathbf{x}_i\right)$ represents the experimental data values and $f\left(\mathbf{x}_i\right)$ the surface fit at each coordinate point in the array $\mathbf{x}_i$. The parameter $\mu\in[0,1]$ controls the degree of smoothing, i.e. the "stiffness" of the marching sheet (*μ=0* produces a linear least-squares fit while *μ=1* produces the "natural" cubic spline interpolant), here $\mu=0.05$ was used. Using the cubic spline formulation and the currently defined tag surface (initially only the start and end points) the coordinates of the next steps inwards (on both fronts) are estimated using extrapolation.

II.C.3.b. Update extrapolated fronts using weighted averaging

The next step is to update the extrapolated fronts using a weighted average of the potential tag voxels that are found on the tag surface and up to one tag thickness offset from the tag surface in both perpendicular directions. The manually determined tag surface start and end locations are also updated to remove possible user bias. The averaging weights are linearly derived from the voxel intensities such that low intensity voxels are favored. These are more likely to be part of tags and this ensures that the surface closely follows the centre of the tags. After this update the next steps are extrapolated until the two moving fronts reach the opposite side of the field of view. The fronts thus cross each other after meeting in the middle and each coordinate is effectively updated twice. This is to remove possible “overshoot” and “undershoot” bias due to the marching direction of the sheet. Once the fronts have reached the opposite side a final surface is fitted to all data points using $\mu = 0.05$. At surface locations where no tag-voxels are found, a cut or void is introduced in the surface. This also allows for handling of local surface shearing. Fig. 4 illustrates the stepwise sheet marching for one of the phantom tag surfaces. Tag surfaces were segmented for both the phantom and volunteer data and for all directions and repetitions (see column 3, Table I). In addition a single average surface set was constructed from the various repetitions for each direction (Fig. 5).

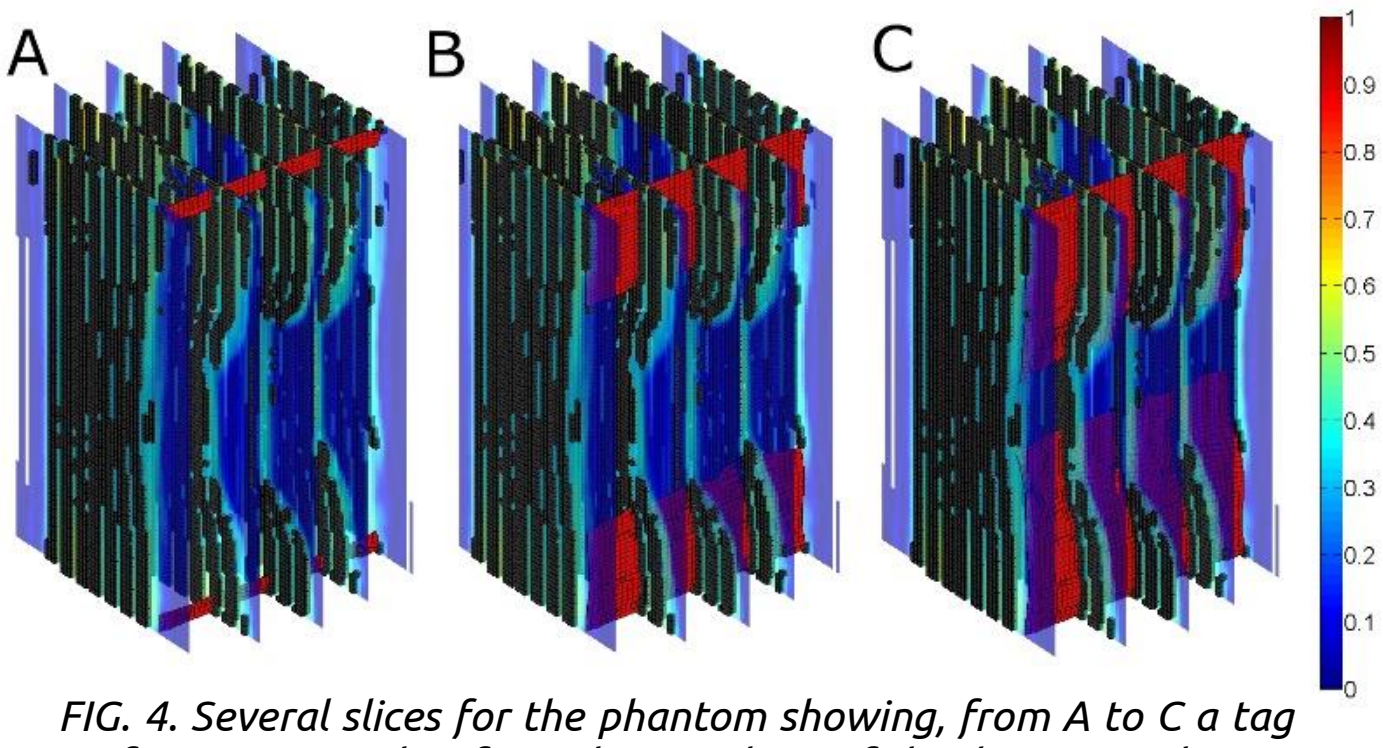

*FIG. 4. Several slices for the phantom showing, from A to C a tag surface as it marches from the periphery of the data inwards. Tag voxels that guide the process are shown in black.*

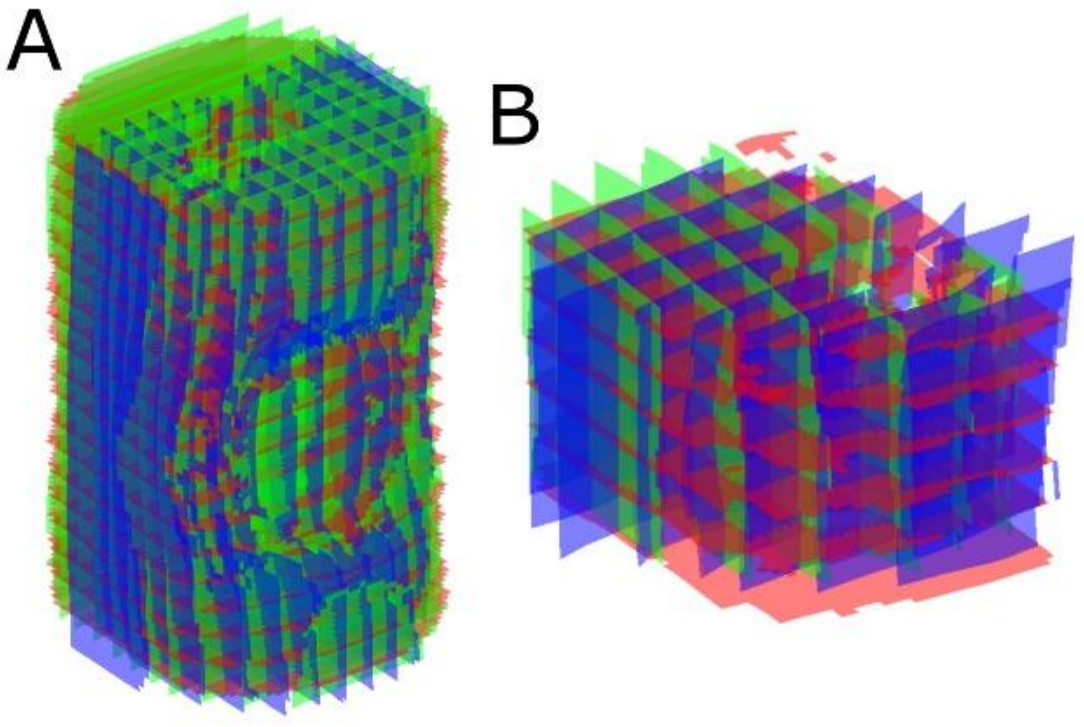

*FIG. 5. The segmented tag surfaces for the phantom (A) and volunteer data (B) in the deformed configuration.*

### *II.C.4. CALCULATION OF TAG SURFACE INTERSECTIONS AND DISPLACEMENT FIELD*

Fig. 5 demonstrates how the average tag surfaces for the three orthogonal directions intersect each other. The next step was to find the intersections for these surfaces as these represent material points that can be tracked over time. This was done as follows: first the shape of a first surface was sampled onto all coordinates defining the second surface. Then the intersection curve of these two surfaces was found by solving where the subtraction of these surfaces equals zero. Finally the intersection point of all three surfaces can then be found by calculation of the intersection of this curve with the third surface. Using this approach all intersection points or tag points for the initial and final configurations could be found. Since tag surfaces were numbered this provided a reference to match corresponding tag points in both configurations. As such the initial and final coordinates could easily be used to construct a 3D displacement vector field, whereby the displacement was simply defined as the difference between the deformed and initial coordinate sets. This approach was followed for all individual surface sets and also for the overall average surface sets. Fig. 6 shows the average vector fields (derived from the average surface sets) obtained for the phantom and volunteer data. The arrow orientations indicates displacement direction and the arrow lengths the magnitude of displacement. For both the phantom and volunteer data the displacement field demonstrates the inhomogeneous nature of the deformation induced by the indentation.

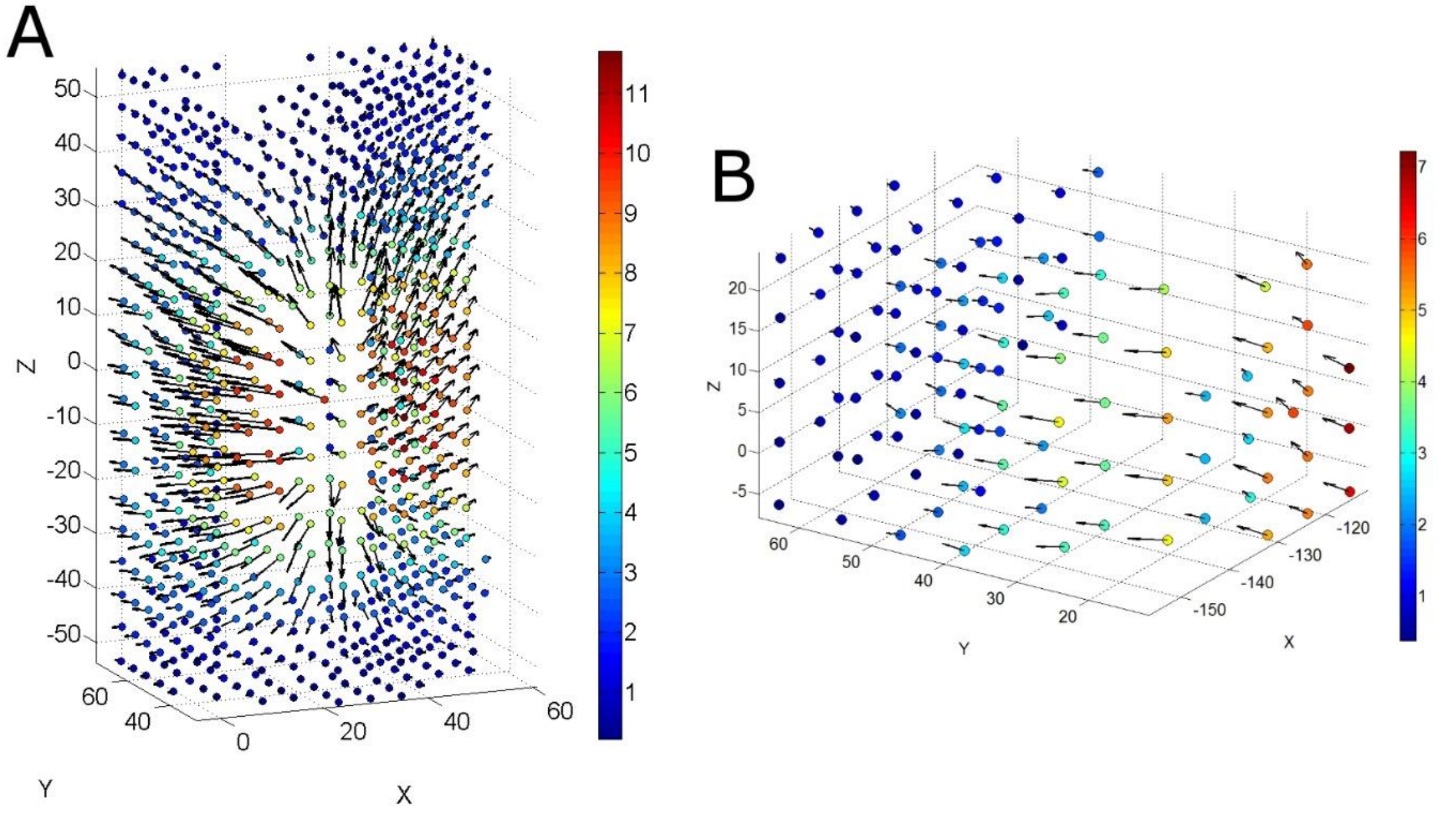

*FIG. 6. The average displacement fields for the phantom (A) and human volunteer data (B). Initial points are coloured according to the displacement magnitude (in mm) and the displacement vector arrows point towards the tag points in the deformed configuration.*

## II.D. Analysis of precision and accuracy

For the current study the following precision and accuracy measures were evaluated: *1) Precision of tag point location, 2) Precision of displacement magnitude* and 3*) Accuracy of displacement measurement in the phantom*.

As shown in column 3 of Table I each tagging direction was repeated several times. This allows for the generation of a large number of combinations of initial and deformed surface sets whose intersections yield the initial and deformed tag point sets. For instance, for the phantom data, there were 700 (10x10x7) possible combinations of initial and 150 (5x5x6) combinations of deformed tag point sets, resulting in 105000 (700x150) possible displacement field combinations (see column 3 of Table I). The repetitions and combinations can be used to analyze the precision of the methods employed. However to limit computational time the precision was analyzed by using 20 random combinations of initial and deformed surface sets which resulted in 20 initial and deformed tag point sets and 400 (20x20) displacement vector fields for both the volunteer and phantom data set. Each coordinate set was compared to the overall average coordinate sets to evaluate precision of tag point locations. Similarly, the precision of the displacement magnitude was assessed by comparison to the overall average displacement magnitude. Finally for the phantom the average displacement field was compared to the "gold standard" displacement measured using marker tracking. This was done by using the average displacement field and the initial marker locations to predict (using interpolation) the marker locations in the deformed configuration. The difference between this predicted displacement and the actual marker displacement provides a measure of the accuracy of the SPAMM tagged MRI based displacement measurement. Gaussian mixture distributions (see Matlab *gmdistribution* function and[43]) were used for statistical analysis of the tag point location precision and the displacement accuracy. The overall mean was defined as the quadratic mean of the means in the X, Y and Z directions while the overall standard deviation was defined as the square root of the mean Eigen-value of the co-variance matrix. In addition root mean square (RMS) values were computed for comparison to values in the literature.

# III. RESULTS

## III.A. Precision of tag point location

As outlined in the Methods section, 20 random tag point sets were evaluated for both the initial and deformed configuration. For each tag point the location difference with respect to the corresponding average tag point was calculated. For the phantom and volunteer data 41040 (2052 tag points/set and 20 random combinations) and 5760 (288 tag points/set and 20 random combinations) tag points were evaluated. Fig. 7 shows scatter plots for all tag point location differences for the phantom (A and B) and volunteer data (C and D). Note that most points are concentrated at the centre and therefore overlap.

For the phantom data the overall mean location difference and standard deviation were 3 μm and 42 μm respectively for the initial configuration. Similarly for the deformed configuration the overall mean and standard deviation were 5 μm and 59 μm. The RMS values were 74 μm and 103 μm for the initial and deformed configuration respectively. The largest tag point location difference in the initial and deformed configurations was 1.41 mm and 1.46 mm respectively. These outliers (differences larger than 250 μm) represented less than 1 % for both the initial and deformed configuration and are always found at the edges and corners of surfaces and surface interruptions (e.g. "bone-like" core) where surface segmentation is based on less information.

Similarly for the volunteer data the overall mean and standard deviation were 3 μm and 118 μm respectively for the initial configuration and 7 μm and 137 μm for the deformed configuration. The RMS values were 204 μm and 238 μm for the initial and deformed configuration respectively. The largest tag point location difference in the initial and deformed configurations was 0.75 mm and 1.27 mm respectively. Again these differences represented isolated cases that were rare in occurrence as differences of magnitudes over 0.5 mm represented only 0.9 % and 1.5 % for the initial and deformed configurations respectively.

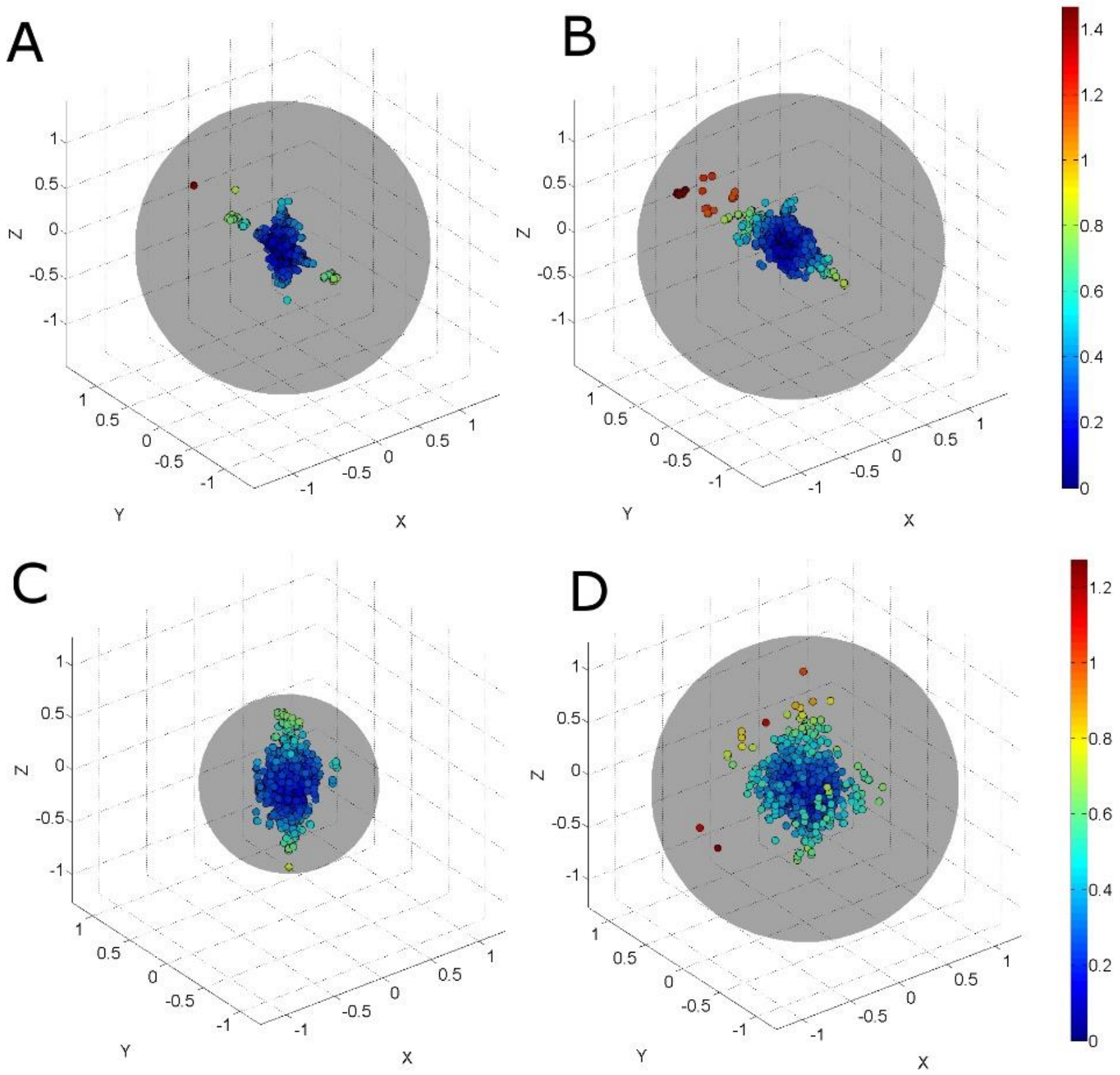

*FIG. 7. Tag point location difference scatter plots for the phantom (A and B) and human volunteer data (C and D) for both the initial (A and C) and deformed configuration (B and D). Color represents differences with respect to average (in mm), outer sphere radius is equal to the maximum difference.*

## III.B. Precision of displacement magnitude

For both the phantom and volunteer data a total of 400 displacement fields were derived from the 20 random tag point sets in the initial and deformed configurations. For each location the difference in displacement magnitude with respect to the corresponding average displacement magnitude (derived from the average surface set intersections see section II.C and IID) was calculated. To study the effect of displacement on the precision of displacement magnitude measurement Fig. 8A shows the distribution of displacement magnitude differences for the phantom data as a function of the average displacement magnitude. Fig. 8B is a plot of the mean and standard deviation of the displacement magnitude differences as a function of the average displacement magnitude of the phantom data. Similarly for the volunteer data this is shown in Fig. 8C and 8D.

For both datasets the mean displacement magnitude differences varied little with increasing displacement. The overall mean and standard deviation for the phantom data was 6 μm and 75 μm respectively and for the volunteer data the mean and standard deviations were 5 μm and 169 μm respectively. However there were some outliers present for the higher displacement levels up to a maximum of 1.48 mm for the phantom and 1.13 mm for the volunteer data. Displacement magnitude differences larger than 0.5 mm represented only 0.1 % and 0.2 % for the phantom and volunteer data respectively. As expected the larger differences occurred in locations where reduced surface information was available in the initial and or deformed configurations (see also section III.A). These were similarly identifiable allowing exclusion if desired. Since the deformation mode was indentation against a bone (or "bone-like" core), the larger displacements also coincide with locations where segmentation of tag surfaces was more challenging and or incomplete, e.g. the gap introduced by the bone. This therefore partially explains the slight increase in variation with displacement magnitude visible in Fig. 8B and 8D.

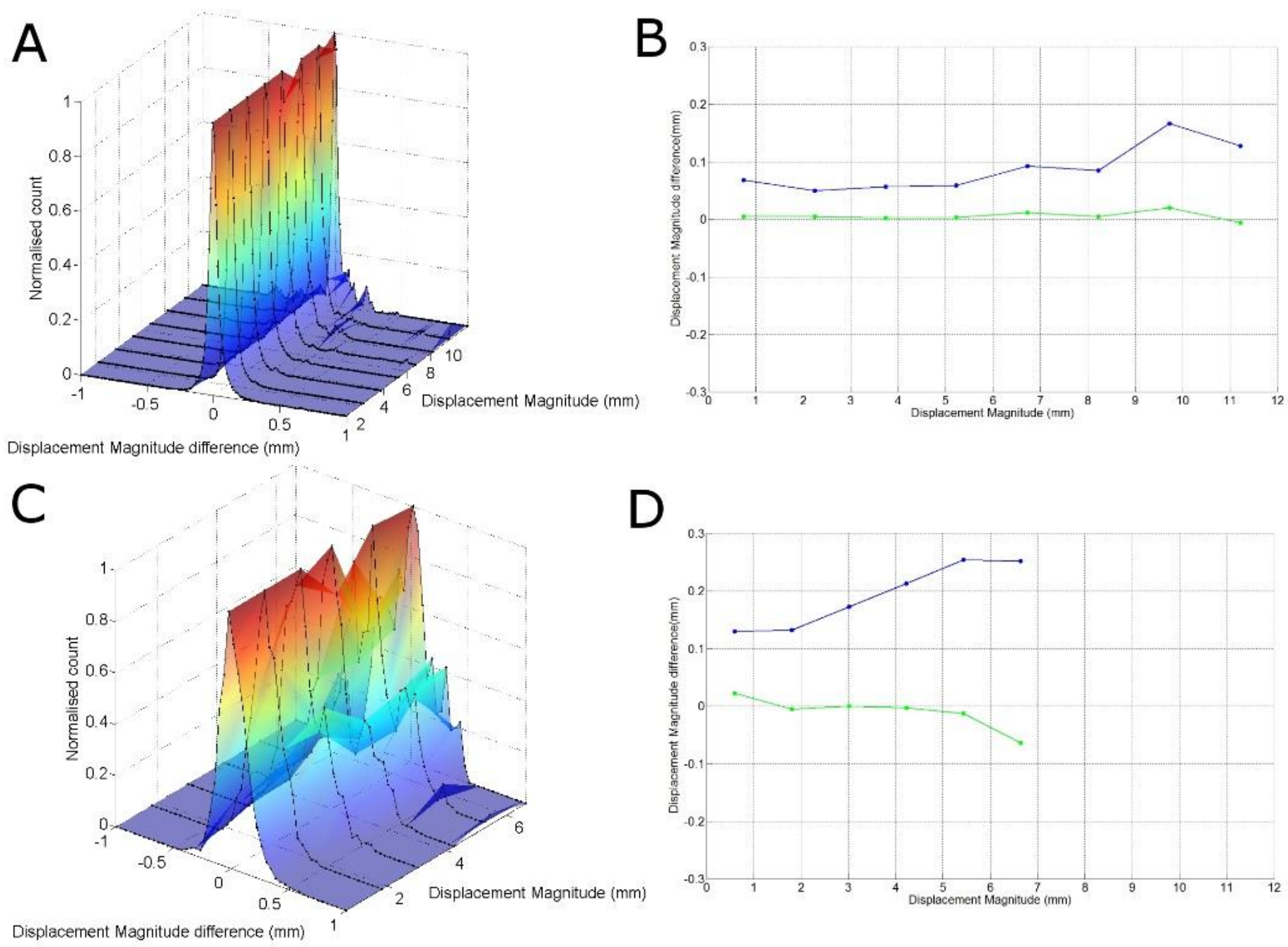

*FIG. 8. The normalised distributions of the displacement magnitude differences as a function of the average displacement magnitude for the phantom (A) and human volunteer data (C), and the mean displacement magnitude difference (green) and standard deviation (blue) as a function of the displacement magnitude for both the phantom (B) and human volunteer data (D).*

## III.C. Accuracy of displacement measurement in the phantom

Fig. 9A shows the marker locations in the initial (blue) and deformed configuration (green) for markers (n=34) that were properly embedded within the (convex hull of the) average displacement field (Fig. 6A). Following comparison with marker displacement, the accuracy of the tag point displacement measurement was evaluated. Using the average displacement field the marker locations in the deformed configuration were predicted (red in Fig. 9A). Note how the predicted and measured marker locations in the deformed configurations overlap. The difference between the predicted marker displacement and the actual marker displacement provides a measure of the accuracy of the displacement measurement. Fig. 9B shows the differences as a 3D scatter plot. The overall mean difference and standard deviation was 72 μm and 289 μm respectively. The RMS values for the displacement differences in the X, Y and Z directions were 253 μm, 354 μm and 278 μm respectively. The maximum difference (891 μm) was found near the edge of the displacement field where predictions were more limited due to the reduced amount of displacement information available here. No relationship with displacement magnitude was observed.

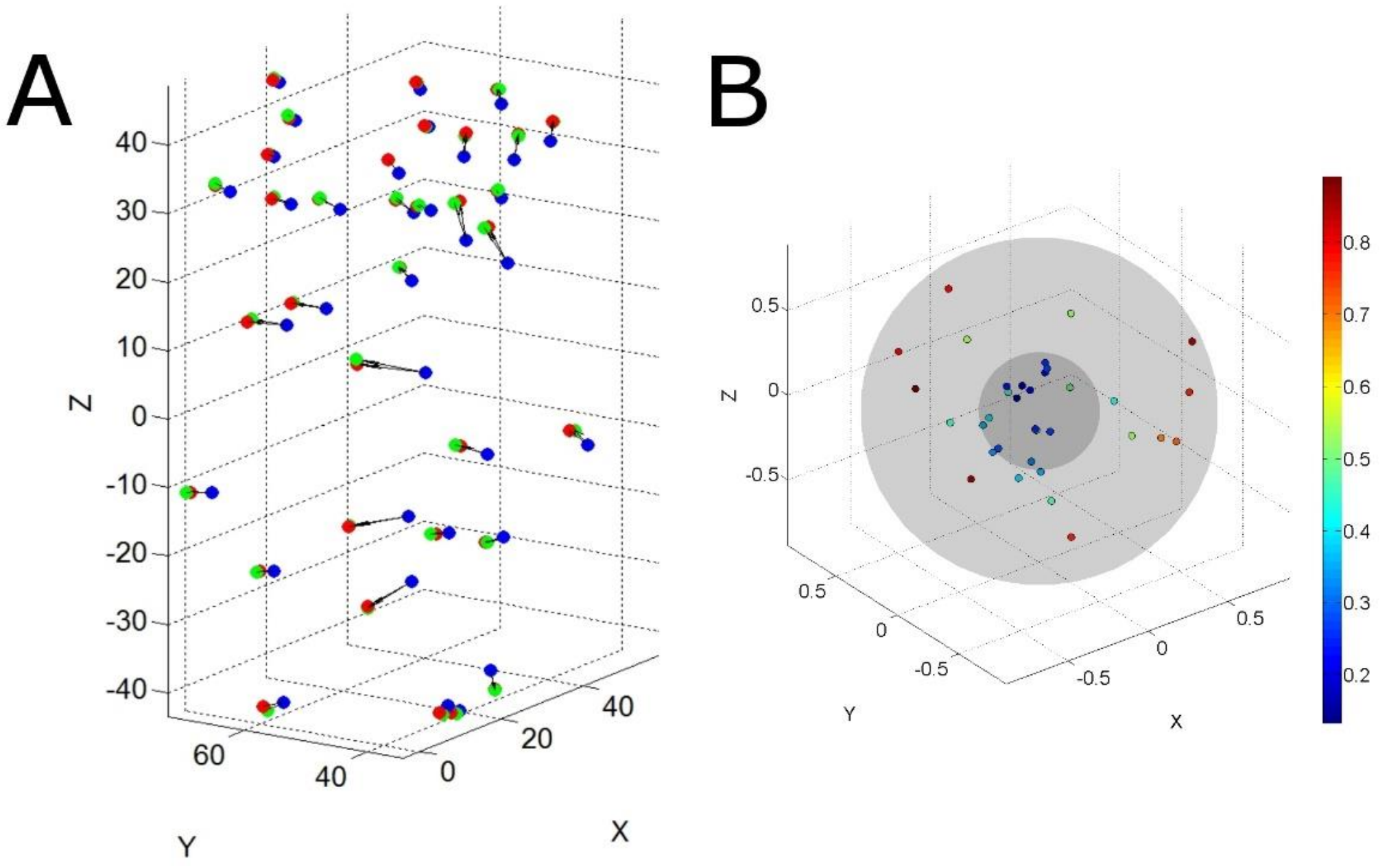

*FIG. 9. The 34 markers in their initial (blue), deformed (green) and predicted deformed locations (red) (A), and a scatter plot for the difference between the measured and predicted marker locations in the deformed configuration. Points are colored towards the magnitude of the difference (mm)*

# IV. DISCUSSION

MRI based measurement of soft tissue motion combined with inverse (finite element) analysis is a powerful method for soft tissue constitutive model parameter identification, which is vital in a diverse range of applications including impact biomechanics[9], rehabilitation engineering[11], surgical simulation[13], soft tissue drug transport[44] and diagnostic medicine e.g. tumour detection[6]. The work presented here is part of a study aiming to use indentation tests on the human arm and the SPAMM tagged MRI based measurement of (complex 3D) soft tissue deformation for the determination of the mechanical properties of passive living human skeletal muscle tissue.

Current SPAMM tagged MRI approaches to non-invasively measure human soft tissue deformation *in-vivo* require the combination of many repeated motion cycles. This has so far generally limited the biomechanical and clinical applications of SPAMM tagged MRI to the study of highly repeatable and periodic movements such as those present in the heart[1, 2, 24]. The large number of repetitions is undesirable for biomechanical applications such as muscle indentation where it causes volunteer discomfort, long scanning times and hinders the analysis of the dynamic biomechanical tissue response. Therefore the purpose of the current study was to develop and validate SPAMM tagged MRI based methods to accurately measure quasi-static 3D soft tissue deformation requiring a minimum of repeated motion (indentation) cycles.

A novel (non-ECG-triggered) SPAMM tagged MRI sequence (3D expansion of [37]) for the measurement of quasi-static 3D soft tissue deformation has been presented. In order to derive 3D soft tissue deformation a total of six (3D TFE) acquisitions are used; three repeated data sets with mutually orthogonal 1-1 (first order) SPAMM directions for both the initial and the deformed configuration. Tag surfaces were segmented using a novel sheet marching algorithm and tag surface. The intersection points of the segmented tag surfaces from three initially orthogonal directions provided trackable material points throughout the volume and allowed for the measurement of 3D deformation. To our knowledge, since the presented methods require the acquisition and combination of only three deformed and three initial configurations the presented methodology is the fastest SPAMM tagged MRI method available for the non-invasive measurement of quasi-static 3D soft tissue deformation.

Quasi-static indentation tests, using an MRI compatible soft tissue indentor, and marker tracking[36] were performed on a silicone gel soft tissue phantom to validate the ability of the proposed methodology to measure 3D soft tissue deformation. The derived displacement demonstrated sub-voxel accuracy with a mean displacement difference of 72 μm and a standard deviation of 289 μm. The performance of the methodology *in-vivo* was also demonstrated using indentation of the biceps region of the upper arm of a volunteer. In addition, several precision measures were evaluated for both the phantom and volunteer data. For the silicone gel phantom the tag point location precision showed a mean and standard deviation of 3 μm and 42 μm respectively for the initial configuration, and 5 μm and 59 μm respectively for the deformed configuration. Similarly for the volunteer, the tag point location precision showed a mean and standard deviation of 3 μm and 118 μm respectively for the initial configuration, and 7 μm and 137 μm respectively for the deformed configuration. In addition displacement magnitude precision was evaluated for both data sets. For the phantom the displacement magnitude precision showed a mean and standard deviation of 6 μm and 75 μm respectively and, similarly for the volunteer, a mean and standard deviation of 5 μm and 169 μm respectively. The sub-voxel accuracy and precision demonstrated in the phantom in combination with the precision comparison between the phantom and volunteer data provide confidence in the methods presented for measurement of soft tissue deformation *in vivo*.

Comparison of the validation results with previous studies is difficult since among other things the deformation modes and magnitudes investigated and the nature of the reference measure vary greatly. However, some comparison with previous methods is appropriate to demonstrate the benefits of the new approach. Young et al. 1993 [45] recorded angular displacement of a silicone gel phantom using tagged MR images and evaluated the results using numerical and analytical modeling and 2D surface deformation derived from optical tracking of lines painted on the phantom surface. A root mean square (RMS) error for the longitudinal translation between magnetic tags and the analytical model (verified by comparison with the optically measured deformation of the painted stripes) was 0.24 mm. Similarly, Chen et al. 2010 [46] recently validated their tagging methods using a numerical phantom and reported RMS errors ranging from 0.15~0.37 mm (depending on cardiac phase). Xu et al. 2010 [47] validated 3D tagging analyzed using optical flow methods using simulated deformation. The displacement in X, Y and Z direction demonstrated RMS errors of 0.43 mm, 0.45 mm and 1mm respectively. For the current study the measurement of deformation was validated using marker tracking in a silicone gel soft tissue phantom demonstrating sub-voxel accuracy with a mean displacement difference of 72 μm and a standard deviation of 289 μm (RMS values for the displacement differences in the X, Y and Z directions were 253 μm, 354 μm and 278 μm respectively). The errors presented in the current study are likely due to noise however despite the fact that only six acquisitions (three static and three indentations) are used, and that SNR levels may be relatively low, the accuracy and precision of the methodology presented are of similar order of magnitude to recent methods involving many repetitions.

Initial tag surface locations are currently determined from three initial configuration data sets. These can be acquired rapidly in series. It is however possible to avoid the use of these initial configurations through certain assumptions about the initial state of tag surfaces. If tag surfaces in the deformed configuration can be segmented successfully and are sufficiently continuous (not sheared), and if their initial state is assumed planar, and if no rigid body movement occurred, then the planar segments of the tag surfaces in un-deformed regions can be used for approximation of initial tag surface shapes and locations. However this may result in inaccuracies as field in-homogeneities are known to result in non planar initial tag surface shapes.

Presently the sheet marching algorithm is seeded using manually determined start and end tag surface locations. However the validation presented here is independent of these manually determined initial estimates since they were updated and replaced by coordinates based on the data. Although the time required for these manual processes is deemed short (~10 minutes of parameter adjustment and 15 minutes of GUI assisted manual start and end point selection) for both current and possibly most clinical applications, in the future it is possible to replace the manual seeding by automatic start and end location identification.

The present study has several limitations. The methods presented here have only been evaluated for measurement of quasi-static deformation measurement, since for each motion cycle the deformation only occurred during the time delay introduced between the pre-pulse and tissue was static during the read-out (and during time-delay for acquisition of initial configurations). Therefore the methods have only been validated and evaluated for quasi-static motions and for cases where the motion is controllable in the sense that three static initial configurations and three repeatable motion cycles can be measured. Such cases include biomechanical applications involving external soft tissue indentation. The performance of the methods presented was not evaluated for dynamic applications whereby motion also occurs during the read-out. In these applications the measured motion would be a combination of that occurring during the delay time and during the read-out time. Despite the quasi-static nature of the methods presented they do enable, when combined with inverse analysis, the investigation of complex material properties such as anisotropy, non-linearity elastic or estimates visco-elasticity. This however requires the measurement of multiple indentation depths and the complex force history boundary conditions.

Future work will focus on the further reduction of the number of repetitions required and on the expansion of these methods to dynamic motion measurements. In addition these methods will be combined with iterative inverse FEA for the determination of the mechanical properties of living human muscle tissue which will improve the predictive capabilities and bio-fidelity of computational models.

# V. CONCLUSIONS

A novel MRI sequence based on SPAMM for the measurement of 3D soft tissue deformation following just six acquisitions (three static and three indentations) has been developed. Its ability to measure soft tissue deformation was validated through indentation tests (using an MRI compatible soft tissue indentor) and marker tracking in a silicone gel phantom. In addition the technique's ability to measure soft tissue deformation *in-vivo* was demonstrated using indentation of the biceps region of the upper arm of a volunteer. Following comparison to marker tracking in the phantom, the SPAMM tagged MRI derived displacement demonstrated sub-voxel accuracy and precision with a mean displacement difference of 72 μm and a standard deviation of 289 μm. Displacement magnitude precision was evaluated for both data sets. The standard deviations of displacement magnitude with respect to the average displacement magnitude were 75 μm and 169 μm for the phantom and volunteer data respectively. The sub-voxel accuracy and precision demonstrated in the phantom in combination with the precision comparison between the phantom and volunteer data provide confidence in the methods presented for measurement of soft tissue deformation *in vivo.* Since only six acquisitions (three static and three indentations) are required the presented methodology is, to our knowledge, the fastest currently available for the non-invasive measurement of 3D soft tissue deformation. This therefore allows for the expansion of the application of SPAMM tagged MRI to (quasi-static) biomechanical applications where a large number of repetitions is undesirable.

# ACKNOWLEDGMENTS

This research was partly funded by Science Foundation Ireland (Research Frontiers Grant 06/RF/ENMO76) and Philips Healthcare. The financial support agencies were not involved in designing and conducting this study, did not have access to the data, and were not involved in data analysis and/or preparation of this manuscript.